\documentstyle[12pt]{article}
 
\begin{document}

\baselineskip=14pt plus 0.2pt minus 0.2pt
\lineskip=14pt plus 0.2pt minus 0.2pt

%***********************
\newcommand{\be}{\begin{equation}}
\newcommand{\ee}{\end{equation}}
\newcommand{\da}{\dagger}
\newcommand{\dg}[1]{\mbox{${#1}^{\dagger}$}}
\newcommand{\hlf}{\mbox{$1\over2$}}
\newcommand{\lfrac}[2]{\mbox{${#1}\over{#2}$}}
%**********************

\begin{flushright}
quant-ph/9608009 \\
LA-UR-96-2756 \\
\end{flushright} 

\begin{center}
\large{\bf  
DISPLACEMENT-OPERATOR SQUEEZED STATES.  \\
II.  EXAMPLES OF TIME-DEPENDENT SYSTEMS \\
HAVING ISOMORPHIC SYMMETRY ALGEBRAS \\}
 
\vspace{0.25in}

\large
\bigskip

Michael Martin Nieto\footnote{\noindent  Email:  
mmn@pion.lanl.gov}\\
{\it Theoretical Division, Los Alamos National Laboratory\\
University of California\\
Los Alamos, New Mexico 87545, U.S.A. \\}
 
\vspace{0.25in}

 D. Rodney Truax\footnote{Email:  truax@acs.ucalgary.ca}\\
{\it Department of Chemistry\\
 University of Calgary\\
Calgary, Alberta T2N 1N4, Canada\\}
 
\normalsize

\vspace{0.3in}

{ABSTRACT}
 
\end{center}
%**************************************
\baselineskip=.33in
%***************************************

\begin{quotation}
In this article, results from the previous paper (I) are applied to 
calculations of squeezed states for such well-known systems 
as the harmonic oscillator, free particle, linear potential, 
oscillator with a uniform driving force, and repulsive 
oscillator.  For each example, expressions for the expectation 
values of position and momentum are derived in terms of the 
initial position and momentum, as well as in the 
$(\alpha,z)$- and in the $(z,\alpha)$-representations described in I.  
The dependence of the squeezed-state uncertainty products 
on the time and on the squeezing parameters are determined for 
each system.
\vspace{0.25in}


\end{quotation}

\newpage

%********************************************************************
\baselineskip=.33in
%******************************************************************
%***********************************

\section{ Introduction}

In paper I, we  discussed the general problem of the squeezed 
states for time-dependent systems in one spatial dimension 
described by the Schr\"odinger equation
\begin{equation}
{\cal S}_1\Psi(x,\tau) = 0,\label{e:I1}
\end{equation}
where the Schr\"odinger operator, ${\cal S}_1$, is 
\begin{equation}
{\cal S}_1 = \partial_{xx} + 2i\partial_{\tau} -2 V(x,\tau).
\label{e:I2}
\end{equation}
The interaction, $V(x,\tau)$, that we considered has the  
form
\begin{equation}
V(x,\tau) = g^{(2)}(\tau)x^2 + g^{(1)}(\tau)x + g^{(0)}(\tau),
\label{e:I3}
\end{equation}
where the coefficients, $g^{(j)}(\tau)$, are differentiable,  
piecewise continuous, but otherwise arbitrary.  The solution space 
of (\ref{e:I1}) was denoted by ${\cal F}_{S_1}$.  We obtained 
the generalized squeezed states for this system and discussed 
their properties.  

However, there are several common, well-known systems subsumed by the  
potential $V(x,\tau)$ in (\ref{e:I3}).  Let ${\bf g}$ be the 3-tuple 
${\bf g} = (g^{(2)}(\tau),g^{(1)}(\tau),g^{(0)}(\tau))$.  Then, for example, 
when 
${\bf g} = (\lfrac{1}{2}\omega^2,0,0)$, we are dealing with the  
simple harmonic oscillator (HO).  If ${\bf g} = (0,0,0)$, then we have a  
free particle (FP).  For the driven harmonic oscillator (DHO), we 
have  ${\bf g} = (\lfrac{1}{2}\omega^2,g(\tau),0)$.   
Two other systems of interest are the linear potential (LP), where  
${\bf g} = (0,g(\tau),0)$, and the repulsive oscillator (RO) for 
which   ${\bf g} =  (-\lfrac{1}{2}\Omega^2,0,0)$.  For both LP and DHO, 
we investigate the specific case $g(\tau)=\kappa/2$, where $\kappa$ is 
a real constant.  As is clear from I,  
all of these systems have isomorphic space-time symmetry algebras 
{\cite{drt1,drt2,gt}}. Then, from our general results, we have  
algebraically calculate 
solution spaces for all these isomorphic systems and hence 
obtain their properties.

In Section 2, for convenience, we present a resum\'e of results from I.  
The time-dependent functions and symmetry operators for each of the 
systems mentioned above are calculated in Section 3.  
Expectation values and uncertainty products for squeezed states 
for each of these examples are presented and discussed in Section 4.  
Finally, we summarize and comment on our results in Section 5. 

%************************

\section{Resum\'e of General Results}

The generators of space-time symmetries were given in Eqs. (6) 
thought (11) in I and will not be repeated here.  The specific nature of 
the $\tau$-dependent solutions of the Schr\"odinger equation were 
determined in Section 3 of I.  We called these solutions ``number-operator 
states."  Some of the real and complex $\tau$-dependent are important here 
and we repeat their definitions below.  Note that the complexified 
Lie algebra $os(1)$ is useful for computing expectation values for 
position and momentum operators.  However, when dealing with specific 
examples, we have found it easier to compute expectation values in terms 
of real $\tau$-dependent functions. 

The function $\xi$ of $\tau$, 
\begin{equation}
\xi(\tau) =\lfrac{1}{\sqrt{2}}(\chi_1(\tau) + i\chi_2(\tau)),
\label{e:II1}
\end{equation}
and its complex conjugate, $\bar{\xi}$, are complex solutions of  
the second order differential equation 
\begin{equation}
\ddot{a} + 2g_2(\tau)a = 0.\label{e:II4}
\end{equation}
The $\tau$-dependent functions, $\chi_1$ and $\chi_2$, are 
real, linearly independent solutions of Eq. (\ref{e:II4}).  The 
properties of these solutions are given in detail in I and 
in Refs. {\cite{drt1,drt2,kt1}}. 

The remaining auxiliary $\tau$-dependent functions of interest in 
this paper are 
\begin{equation}
{\cal C}(\tau) = c(\tau)+{\cal C}^o, \label{e:II10c}
\end{equation}
where ${\cal C}^o$ is a complex integration constant and $c(\tau)$ 
is the definite integral
\begin{equation}
c(\tau) = \int_{0}^{\tau}d\rho\,\xi(\rho)g^{(1)}(\rho)\label{e:II11c}
\end{equation}
and
\begin{equation}
\phi_1(\tau) = \xi^2,~~\phi_2(\tau) = \bar{\xi}^2,~~
\phi_3(\tau) = 2\xi\bar{\xi}, \label{e:II14c}
\end{equation}

The complex function ${\cal C}$ can be written {\cite{kt1}} 
in terms of real functions ${\cal C}_1$ and ${\cal C}_2$ 
\begin{equation}
{\cal C}(\tau) = \lfrac{1}{\sqrt{2}}\left({\cal C}_1(\tau)
+i{\cal C}_2(\tau)\right), \label{e:II22}
\end{equation}
where
\begin{equation}
{\cal C}_{\nu}(\tau) = c_{\nu}(\tau)+{\cal C}_{\nu}^o
\label{e:II10r}
\end{equation}
and
\begin{equation}
c_{\nu}(\tau) = \int_{0}^{\tau}d\rho\,\chi_{\nu}(\rho)g^{(1)}(\rho),
\label{e:II11r}
\end{equation}
for $\nu = 1,2$.  The complex integration constant ${\cal C}^o$ and 
its complex conjugate are related to the real integration constants, 
${\cal C}_1^o$ and ${\cal C}_2^o$ by
\begin{equation}
{\cal C}^o = \lfrac{1}{\sqrt{2}}\left({\cal C}_1^o+i{\cal C}_2^o\right). 
\label{e:II22o}
\end{equation}
Also, we define the following initial values for the real and 
complex functions:
\begin{eqnarray}
 & \xi^o = \xi(0),~~~\bar{\xi}^o=\bar{\xi}(0),~~~
\phi_3^o= \phi_3(0), & \label{e:II25c}\\*[1.5mm]
 & \chi_1^o=\chi_1(0),~~\chi_2^o=\chi_2(0),~~
\hat{\phi}_3^o = \hat{\phi}_3(0).\label{e:II25r}
\end{eqnarray}

The coherent and squeezed state parameters are,
\be
\alpha = |\alpha|e^{i\delta},~~z = re^{i\theta},~~r=|z|.
\label{e:II30}
\ee
The expectation values of position and momentum in terms of  
complex functions and initial position and momentum are given 
by
\begin{eqnarray}
\langle x(\tau)\rangle & = & 
i\left\{[\bar{\xi}\xi^o-\xi\bar{\xi}^o]p_o
-[\xi\dot{\bar{\xi}}^o-\bar{\xi}\dot{\xi}^o]x_o\right\}\nonumber\\
                       &   & 
~~~~~+i\left(\xi\bar{c}- \bar{\xi}c\right),
\label{e:II35c}\\[2mm]
\langle p(\tau)\rangle & = & 
i\left\{[\dot{\bar{\xi}}\xi^o - \dot{\xi}\bar{\xi}^o]p_o  
+[\dot{\xi}\dot{\bar{\xi}}^o -\dot{\bar{\xi}}\dot{\xi}^o]\right\}
\nonumber\\*[1.5mm]
                       &   & 
~~~~~+i\left(\dot{\xi}\bar{c} - \dot{\bar{\xi}}c\right),
\label{e:II40c}
\end{eqnarray}
where for the $(\alpha,z)$-representation 
\begin{equation}
\alpha = i\left(p_o \xi^o - x_o \dot{\xi}^o\right)+i{\cal C}^o,
\label{e:II45c}
\end{equation}
and in the $(z,\alpha)$-representation,
\begin{equation}
|\alpha|[e^{i\delta}\cosh{r} - e^{i(\theta-\delta)}\sinh{r}] =  
i\left(p_o\xi^o-x_o\dot{\xi}^o\right)+i{\cal C}^o.\label{e:II46c}
\end{equation}
Similarly for  their complex conjugates.
 
Now we write the expectation values of position and momentum  
explicitly in terms of real $\tau$-dependent functions.  These are 
more useful in constructing properties of squeezed states for 
specific systems.  The expectation value for position is 
\begin{eqnarray}
\langle x(\tau)\rangle & = & 
\left(\chi_2\chi_1^o+\chi_1\chi_2^o\right)p_o 
-\left(\chi_1\dot{\chi}_2^o-\chi_2\dot{\chi}_1^o\right)x_o
\nonumber\\
                       &   & 
+\chi_1c_2-\chi_2c_1, \label{e:II35r}
\end{eqnarray}
and for momentum is
\begin{eqnarray}
\langle p(\tau)\rangle & = & 
\left(\dot{\chi}_2\chi_1^o+\dot{\chi}_1\chi_2^o\right)p_o 
-\left(\dot{\chi}_1\dot{\chi}_2^o-\dot{\chi}_2\dot{\chi}_1^o\right)
x_o\nonumber\\
                       &   & 
+\dot{\chi}_1c_2-\dot{\chi}_2c_1.\label{e:II40r}
\end{eqnarray}
The coherent and squeezed state parameters can also be expressed in terms 
of the initial position and momentum and the initial values of the real 
$\tau$-dependent functions.  For the $(\alpha,z)$-representation, the 
equations are
\begin{eqnarray}
|\alpha|^2 & = & 
\lfrac{1}{2}\left[\left(\dot{\chi}_2^ox_o-\chi_2^op_o
-{\cal C}^o_2\right)^2
+\left(\chi_1^op_o-\dot{\chi}_1^ox_o+{\cal C}^o_1\right)^2\right],
\nonumber\\
\tan{\delta} & = & {{\chi_1^op_o-\dot{\chi}_1^ox_o+{\cal C}_1^o}
\over{\dot{\chi}_2^ox_o-\chi_2^op_o-{\cal C}_2^o}}.\label{e:II45r}
\end{eqnarray}
For the $(z,\alpha)$-representation, we have
\begin{eqnarray}
|\alpha|\left[\cos{\delta}\cosh{r}-\cos{(\theta-\delta)}\sinh{r}\right]
 & = & \sqrt{\lfrac{1}{2}}\left(\dot{\chi}_2^ox_o-\chi_2^op_o
-{\cal C}_2^o\right),\nonumber\\
|\alpha|\left[\sin{\delta}\cosh{r}-\sin{(\theta-\delta)}\sinh{r}\right]
 & = & \sqrt{\lfrac{1}{2}}\left(\chi_1^op_o-\dot{\chi}_1^ox_o
+{\cal C}_1^o\right).\label{e:II46r}
\end{eqnarray}

Frequently, we wish to use the expressions for the expectation values 
of position and momentum in terms of the parameters $\alpha$ and $z$.  
We give both the $(\alpha,z)$- and $(z,\alpha)$-representations in terms 
of real $\tau$-dependent functions only.  The complex expressions can be 
found in paper I. 

\vspace{.5cm}
\noindent {\it The} $(\alpha,z)${\it -representation}

Substituting Eqs.(\ref{e:II1}) and (\ref{e:II22}) into Eqs. 
(73) and (74) from paper I, we calculate real expressions for 
$\langle x(\tau)\rangle_{(\alpha,z)}$ and 
$\langle p(\tau)\rangle_{(\alpha,z)}$: 
\begin{equation}
\langle x(\tau)\rangle_{(\alpha,z)} = \sqrt{2}|\alpha|
\left(\chi_1\cos{\delta}+\chi_2\sin{\delta}\right)
+\chi_1{\cal C}_2-\chi_2{\cal C}_1\label{e:II50r}
\end{equation}
and
\begin{equation}
\langle p(\tau)\rangle_{(\alpha,z)} = \sqrt{2}|\alpha|
\left(\dot{\chi}_1\cos{\delta}+\dot{\chi}_2\sin{\delta}\right)
+\dot{\chi}_1{\cal C}_2-\dot{\chi}_2{\cal C}_1.
\label{e:II53r}
\end{equation}
\vspace{.5cm}

\noindent {\it The} $(z,\alpha)${\it -representation}

Substituting Eqs. (\ref{e:II1}) and (\ref{e:II22}) into Eqs. 
(79) and (81) from paper I, we obtain the following 
expressions for $\langle x(\tau)\rangle_{(z,\alpha)}$ and 
$\langle p(\tau)\rangle_{(z,\alpha)}$: 
\begin{eqnarray}
\langle x(\tau)\rangle_{(z,\alpha)} & = & \sqrt{2}|\alpha|
\left\{\chi_1\left[\cos{\delta}\cosh{r}-\cos{(\theta-\delta)}
\sinh{r}\right]\right.\nonumber\\
                                    &   & 
+\left.\chi_2\left[\sin{\delta}\cosh{r}-\sin{(\theta-\delta)}
\sinh{r}\right]\right\}\nonumber\\
                                    &   & 
+\chi_1{\cal C}_2-\chi_2{\cal C}_1,\label{e:II55r}
\end{eqnarray}
\begin{eqnarray}
\langle p(\tau)\rangle_{(z,\alpha)} & = & \sqrt{2}|\alpha|
\left\{\dot{\chi}_1\left[\cos{\delta}\cosh{r}
-\cos{(\theta-\delta)}\sinh{r}\right]\right.\nonumber\\
                                    &   & 
+\left.\dot{\chi}_2\left[\sin{\delta}\cosh{r}-\sin{(\theta-\delta)}
\sinh{r}\right]\right\}\nonumber\\
                                    &   & 
+\dot{\chi}_1{\cal C}_2-\dot{\chi}_2{\cal C}_1,\label{e:II58r}
\end{eqnarray}
\vspace{.3cm}

The uncertainty product is representation independent and its 
expression in terms of the real $\tau$-dependent functions,
$\chi_1$ and $\chi_2$, is given by Eq. (93) in paper I.  
\vspace{.5cm}

%***********************************

\section{Examples}

\subsection{ Harmonic oscillator (HO)}

For the harmonic oscillator  
${\bf g} = (\lfrac{1}{2}\omega^2,0,0)$ and the differential 
equation (\ref{e:II4}) has the form 
\begin{equation}
\ddot{a} + \omega^2 a = 0.\label{e:ho1}
\end{equation}
Two real solutions are
\begin{equation}
\chi_1 = \lfrac{1}{\sqrt{\omega}}\cos{\omega\tau},
~~~~~\chi_2 =  
\frac{1}{\sqrt{\omega}}\sin{\omega\tau}.\label{e:ho4}
\end{equation}
Using Eq. (\ref{e:II1}), the two complex solutions are
\begin{equation}
\xi = \sqrt{\lfrac{1}{2\omega}}{\rm e}^{i\omega\tau},
~~~~~\bar{\xi} = \sqrt{\lfrac{1}{2\omega}}{\rm
e}^{-i\omega\tau},\label{e:ho6} 
\end{equation}
from which we obtain the three functions $\phi_1$, $\phi_2$, and  
$\phi_3$:
\begin{equation}
\phi_1 = \lfrac{1}{2\omega}{\rm e}^{2i\omega\tau},
~~~~\phi_2 = \lfrac{1}{2\omega}{\rm  
e}^{-2i\omega\tau},~~~~\phi_3 =
\lfrac{1}{\omega}.\label{e:ho10} 
\end{equation}

Therefore, the generators in Eqs. (6) to (11) in paper I have 
the form
\begin{eqnarray}
 & {\cal J}_- = \sqrt{\lfrac{1}{2\omega}}
e^{i\omega\tau}(\partial_x+\omega x),~~~{\cal J}_+ =
\sqrt{\lfrac{1}{2\omega}}e^{-i\omega\tau}(-\partial_x+
\omega x), & \label{e:ho14}
\\*[1.5mm]
 & {\cal M}_- = 
\lfrac{1}{2\omega}e^{2i\omega\tau}(i\partial_{\tau} 
-\omega x\partial_x
-\omega^2 x^2 -\lfrac{1}{2}\omega), & \label{e:ho16}
\\*[1.5mm]
 & {\cal M}_+ =  
\lfrac{1}{2\omega}e^{-2i\omega\tau}(i\partial_{\tau}+
\omega x\partial_x -\omega^2 x^2 +\lfrac{1}{2}\omega), &
\label{e:ho18}
\\*[1.5mm] 
 & {\cal M}_3 = \lfrac{i}{\omega}\partial_{\tau}.\label{e:ho20}
\end{eqnarray}
\vskip .2cm

\subsection{Free particle (FP)}

In this specific case, ${\bf g} = (0,0,0)$ and Eq. 
(\ref{e:II4}) becomes
\begin{equation}
\ddot{a} = 0,\label{e:fp1}
\end{equation}
which has real solutions,
\begin{equation}
\chi_1 = 1,~~~~~~~~\chi_2 = \tau.\label{e:fp4}
\end{equation}
The two complex solutions can be calculated directly using Eq.  
(\ref{e:II1}) 
and they are
\begin{equation}
\xi = \sqrt{\lfrac{1}{2}}\left(1 + i\tau\right),
~~~~~~~~\bar{\xi} = \sqrt{\lfrac{1}{2}}\left(1 - i\tau\right).
\label{e:fp8}
\end{equation}
The remaining $\tau$-dependent functions can be obtained from  
Eqs. (\ref{e:II14c}):
\begin{equation}
\phi_1 = \lfrac{1}{2}\left(1 + i\tau\right)^2,
~~~~~\phi_2 = \lfrac{1}{2}\left(1-i\tau\right)^2,~~~~~\phi_3 =
1+\tau^2.\label{e:fp12} 
\end{equation}

Therefore, the generators in Eqs. (6) to (11) of paper I can be 
written as
\begin{eqnarray}
 & {\cal J}_- = \sqrt{\lfrac{1}{2}}\{\left(1+i\tau\right)
\partial_x + x\},~~~{\cal J}_+ = \sqrt{\lfrac{1}{2}}
\{-\left(1-i\tau\right)\partial_x + x\},  &
\label{e:fp16}
\\*[1mm]
 & {\cal M}_- = \lfrac{i}{2}\left\{\left(1+i\tau\right)^2
\partial_{\tau}+i\left(1+i\tau\right)x\partial_x 
+\lfrac{i}{2}x^2+\lfrac{i}{2}\left(1+i\tau\right)\right\}, & 
\label{e:fp18}
\\*[1mm]
 & {\cal M}_+ = \lfrac{1}{2}\left\{\left(1-i\tau\right)^2
\partial_{\tau}-i\left(1-i\tau\right)x\partial_x 
+\lfrac{i}{2}x^2-\lfrac{i}{2}\left(1-i\tau\right)\right\}, & 
\label{e:fp20}
\\*[1.5mm]
 & {\cal M}_3 = \lfrac{i}{2}\left\{\left(1+\tau^2\right)
\partial_{\tau}+ \tau x\partial_x - \lfrac{i}{2}x^2 +  
\lfrac{1}{2}\tau\right\}. & \label{e:fp22} 
\end{eqnarray}
\vskip .2cm
We have presented the generators in this and the previous example out 
of general interest.  We will not show them in subsequent examples 
since they are usually long and not needed.  In all cases 
described by the potential in Eq. (\ref{e:I3}), the 
calculations for which the generators are required can all be done 
in general.  (See Section 3 of paper I.)  All wave functions and 
expectation values can be obtained by computing the appropriate 
$\tau$-dependent functions and substituting them into expressions 
for the desired quantities.  This is where the strength of this 
methodology lies.

\subsection{Linear potential (LP)}

Here we have ${\bf g}=(0,g(\tau),0)$ and the differential equation  
is (\ref{e:fp1}).  The two real solutions are (\ref{e:fp4}) and  
the two complex solutions are (\ref{e:fp8}).  Set $g(0)=g_o$. 

The $\tau$-dependent function, ${\cal C}$ is defined in Eq. 
(\ref{e:II10r}), where $c(\tau)$ is
\begin{equation}
c(\tau) = \sqrt{\lfrac{1}{2}}\int_0^{\tau}d\rho\,g(\rho)(1+i\rho).
\label{e:lp1}
\end{equation}
The complex conjugate, $\bar{c}(\tau)$, can be obtained directly 
from Eq. (\ref{e:lp1}).  The two generators ${\cal J}_-$ and 
${\cal J}_+$ can be written as
\begin{eqnarray}
{\cal J}_- & = & \sqrt{\lfrac{1}{2}}\left\{\left(1+i\tau\right)
\partial_x+x-i{\cal C}\right\},\nonumber\\*[1mm]
{\cal J}_+ & = & \sqrt{\lfrac{1}{2}}\left\{-\left(1-i\tau\right)
\partial_x+x+i\bar{\cal C}\right\}.\label{e:lp4}
\end{eqnarray}
The commutator of ${\cal J}_-$ and ${\cal J}_+$ is 
\begin{equation}
[{\cal J}_-,{\cal J}_+] = I.\label{e:lp5}
\end{equation}

To determine the integration constants, ${\cal C}^o$ and 
$\bar{\cal C}^o$, calculate the operator 
${\cal J}_+{\cal J}_-+\lfrac{1}{2}$ and find that 
\begin{eqnarray}
{\cal J}_+{\cal J}_-+\lfrac{1}{2} & = & \lfrac{1}{2}
\left\{-(1+\tau^2)\partial_{xx} +2i\tau x\partial_x\right.
\nonumber\\
  &   & \left.+\sqrt{2}\left[i(1+i\tau)\bar{\cal C}+
i(1-i\tau){\cal C}\right]\partial_x+ x^2\right.\nonumber\\
  &   & \left. +i\sqrt{2}(\bar{\cal C}-{\cal C})x+i\tau
+{\cal C}\bar{\cal C}\right\}.\label{e:lp8}
\end{eqnarray}
Taking the limit as $\tau\rightarrow 0$, we have
\begin{equation}
\lim_{\tau\rightarrow 0}\left\{{\cal J}_+{\cal J}_-+\lfrac{1}{2}
\right\} =
\lfrac{1}{2}
\left\{-\partial_{xx}+i\sqrt{2}\left(\bar{\cal C}^o+{\cal C}^o\right)
\partial_x
+x^2+i\sqrt{2}(\bar{\cal C}^o-{\cal C}^o)x+{\cal C}^o\bar{\cal C}^o
\right\},\label{e:lp12}
\end{equation}
which is a Hamiltonian for a displaced oscillator if we 
choose ${\cal C}^o$ to be pure imaginary and set 
\begin{equation}
{\cal C}^o = \lfrac{i}{\sqrt{2}}g_o.\label{e:lp16}
\end{equation}
Therefore, we find that 
\begin{equation}
\lim_{\tau\rightarrow 0}\left\{{\cal J}_+{\cal J}_-+\lfrac{1}{2}
\right\} =
\lfrac{1}{2}\left\{-\partial_{xx}
+x^2+i\sqrt{2}g_ox+\lfrac{1}{2}g_o^2
\right\}.\label{e:lp13}
\end{equation}
Then, we see that 
\begin{equation}
{\cal C}=c(\tau)+\lfrac{i}{\sqrt{2}}g_o, ~~~~~~\bar{\cal C}=
\bar{c}(\tau)-\lfrac{i}{\sqrt{2}}g_o.\label{e:lp20}
\end{equation}
Once $g(\tau)$ is known, the remaining $\tau$-dependent functions 
can be found and the symmetry generators constructed from 
Eqs. (6) to (11) in paper I.  

For example, suppose that we choose 
\begin{equation}
g(\tau)=\frac{\kappa}{2},\label{e:lp24}
\end{equation}
where $\kappa$ is a real constant.  Then, $g_o=\kappa/2$, and 
we have for the complex function ${\cal C}$ 
\begin{equation}
{\cal C}(\tau) = \lfrac{\kappa}{4\sqrt{2}}\left[2\tau+i(2+\tau^2)
\right].\label{e:lp28}
\end{equation}
\vskip .2cm

We will need the real functions ${\cal C}_1$ and ${\cal C}_2$, as 
well as the values of the real integration constants 
${\cal C}^o_1$ and ${\cal C}_2^o$.  They can be evaluated with 
the help of Eqs. (\ref{e:II22}) through (\ref{e:II22o}).  
For arbitrary $g(\tau)$, we have 
\begin{eqnarray}
 & {\cal C}_1(\tau) = c_1(\tau) = \int_0^{\tau}d\rho\,g(\rho), &
\label{e:lp36}\\*[1mm]
 & {\cal C}_2(\tau) = c_2(\tau)+{\cal C}_2^0, & \label{e:lp37} 
\end{eqnarray}
where 
\begin{equation}
c_2(\tau) = \int_0^{\tau}d\rho\,g(\rho)\rho,\label{e:lp38}
\end{equation}
and 
\begin{equation}
{\cal C}_1^o=0,~~~~~~{\cal C}_2^o = g_o.\label{e:lp39}
\end{equation}
When $g(\tau)$ is a constant, $\kappa/2$,  then 
\begin{eqnarray}
 & {\cal C}_1(\tau) = c_1(\tau) = \frac{\kappa}{2}\tau, &
\label{e:lp42}\\*[1mm]
 & {\cal C}_2(\tau) = c_2(\tau)+{\cal C}_2^0, & \label{e:lp43} 
\end{eqnarray}
where
\begin{equation}
c_2(\tau) = \frac{\kappa}{4}\tau^2,\label{e:lp44}
\end{equation}
and 
\begin{equation}
{\cal C}_2^o=\frac{\kappa}{2}.\label{e:lp45}
\end{equation}
\vskip .2cm

\subsection{Driven Harmonic Oscillator (DHO)}

For this system, we have ${\bf g} = (\lfrac{1}{2}\omega^2,  
g(\tau),0)$, where $g(\tau)$ is a real function of $\tau$ and 
$g(0)=g_o$.  The differential equation for $\chi_1$ and 
$\chi_2$ is (\ref{e:ho1}).   The real solutions are given 
in (\ref{e:ho4}) and the complex solutions in  (\ref{e:ho6}).  
The function, ${\cal C}$, can be written as in Eq. 
(\ref{e:II10c}) with
\begin{equation}
c(\tau)=  
\sqrt{\lfrac{1}{2\omega}}\int_{0}^{\tau} d\rho\,g(\rho){\rm
e}^{i\omega\rho}. \label{e:do1}
\end{equation}
From Eq. (\ref{e:do1}) and its complex conjugate all the remaining 
$\tau$-dependent functions can be found, if needed.  

The two operators ${\cal J}_{\pm}$ are
\begin{eqnarray}
{\cal J}_- & = & \sqrt{\lfrac{1}{2\omega}}{\rm e}^{i\omega\tau}
\left\{\partial_x +\omega x +i\sqrt{2\omega}{\rm e}^{-i\omega\tau}
{\cal C}\right\}, \label{e:do4}\\*[1mm]
{\cal J}_+ & = & -\sqrt{\lfrac{1}{2\omega}}{\rm e}^{-i\omega\tau}
\left\{\partial_x -\omega x +i\sqrt{2\omega}{\rm e}^{+i\omega\tau}
{\cal C}\right\}. \label{e:do5}
\end{eqnarray}
Repeating the procedure of the previous subsection, computing the 
operator ${\cal J}_+{\cal J}_-+\lfrac{1}{2}$, and taking the limit 
$\tau\rightarrow 0$, we obtain the integration constant 
\begin{equation}
{\cal C}^o = -\lfrac{i}{\omega\sqrt{2\omega}}g_o,\label{e:do8}
\end{equation}
and its complex conjugate.  

For the specific case when 
\begin{equation}
g(\tau)=g_o=\frac{\kappa}{2},\label{e:do12}
\end{equation}
$\kappa$ a real number, we have 
\begin{equation}
{\cal C} = -\lfrac{i\kappa}{(2\omega)^{3/2}}{\rm e}^{i\omega\tau}.
\label{e:do16}
\end{equation}
\vskip .2cm

The real functions ${\cal C}_1$ and ${\cal C}_2$ and have the 
form Eq. (\ref{e:II10r}) where
\begin{equation}
c_1(\tau) = \int_0^{\tau}d\rho\,g(\rho)\cos{\omega\rho},~~~~
c_2(\tau) = \int_0^{\tau}d\rho\,g(\rho)\sin{\omega\rho},
\label{e:do24}
\end{equation}
with real integration constants 
\begin{equation}
{\cal C}_1^o=0, ~~~~~{\cal C}_2^o = -\lfrac {1}{\omega^{3/2}}g_o.
\label{e:do26}
\end{equation}
When $g(\tau)=\kappa/2$, Eqs. (\ref{e:do24}) become 
\begin{equation}
c_1(\tau)=\lfrac{\kappa}{2\omega^{3/2}}\sin{\omega\tau},
~~~~~c_2(\tau)=-\lfrac{\kappa}{2\omega^{3/2}}
\left(1-\cos{\omega\tau}\right),\label{e:do28}
\end{equation}
with real integration constants  
\begin{equation}
{\cal C}_1^o=0, ~~~~~{\cal C}_2^o = -\lfrac {1}{2\omega^{3/2}}\kappa.
\label{e:do30}
\end{equation}
\vskip .2cm
 
\subsection{Repulsive oscillator (RO)}

In this final example we have 
${\bf g} = (-\lfrac{1}{2}\Omega^2,0,0)$.   Eq. (\ref{e:II4})  
becomes 
\begin{equation}
\ddot{a} - \Omega^2 a = 0.\label{e:ro1}
\end{equation}
This differential equation has two real solutions:
\begin{equation}
\chi_1 = \lfrac{1}{\sqrt{\Omega}}\cosh{\Omega\tau},
~~~~~\chi_2=\lfrac{1}{\sqrt{\Omega}}\sinh{\Omega\tau}.\label 
{e:ro4}
\end{equation}
Using Eq. (\ref{e:II1}), the two complex solutions are
\begin{eqnarray}
\xi & = & \lfrac{1}{\sqrt{2\Omega}}\left[\cosh{\Omega\tau}
+i\sinh{\Omega\tau}\right],\nonumber\\*[1mm]
\bar{\xi} & = & 
\lfrac{1}{\sqrt{2\Omega}}[\cosh{\Omega\tau}-
i\sinh{\Omega\tau}].\label{e:ro8}
\end{eqnarray}
\vskip .4cm

For each of model systems discussed above, 
their Schr\"odinger algebras are isomorphic, 
$({\cal SA})_1^c = su(1,1)\diamond w_1^c.$  
An explicit operator basis for each system can be obtained by 
substituting the appropriate functions for each example into 
Eqs. (6) to (11) of paper I.  As examples, see Eqs. 
(\ref{e:ho14}) to (\ref{e:ho20}) for the harmonic oscillator 
and Eqs. (\ref{e:fp16}) to (\ref{e:fp22}) for the free particle.

In each case, the subalgebra of operators 
$\{{\cal M}_3,{\cal J}_{\pm},I\}$ forms a basis 
for an oscillator algebra $os(1)$.  Therefore, each of the  
examples above will have a complete set of number-operator 
states.  Only, for HO  will these states be energy eigenstates  
since, in that case, ${\cal M}_3$ is proportional to the energy 
operator.  For the remaining examples, there may be no simple  
interpretation for the operator ${\cal M}_3$.

%************************

\section{ Squeezed States for Specific Examples}

Explicit calculation of coherent states for HO, FP, DHO, LP, 
and RO have been calculated elsewhere {\cite{gt}}.  These results 
are contained as specific examples, $z = 0$, of our results here.   

The specific $\tau$-dependent functions that we need were 
developed in Section 3.  In each case, we express the expectation 
values for position and momentum in three ways: in terms of the 
initial position and momentum [Eqs. (\ref{e:II35r}) and 
(\ref{e:II40r})], in the $(\alpha,z)$-representation [Eqs. 
(\ref{e:II50r}) and (\ref{e:II53r})], and in the 
$(z,\alpha)$-representation [Eqs. (\ref{e:II55r}) and 
(\ref{e:II58r})].  However, the parameters $|\alpha|$, 
$\delta$, $|z|$, and $\theta$ are defined in terms of 
$x_o$ and $p_o$ differently in the $(\alpha,z)$- and 
$(z,\alpha)$-representations, Eqs. (\ref{e:II45r}) and 
(\ref{e:II46r}), respectively.  Finally, for each system  
we give the uncertainty product, which is independent 
of representation.   

\subsection{Harmonic oscillator (HO)}

Combining Eqs. (\ref{e:II35r}) and (\ref{e:II40r}) with the 
real functions (\ref{e:ho4}) for the oscillator, we find that 
the expectation values for position and momentum in terms of  
$x_o$ and $p_o$, are
\begin{eqnarray}
\langle x(\tau)\rangle & = &  
\lfrac{1}{\omega}\left(p_o\sin{\omega\tau} +
\omega x_o\cos{\omega\tau}\right),\label{e:ssho1}
\\*[1mm]
\langle p(\tau)\rangle & = & p_o\cos{\omega\tau} -
\omega x_o\sin{\omega\tau}.\label{e:ssho2}
\end{eqnarray}

In the $(\alpha,z)$-representation, using Eqs. (\ref{e:II50r}) 
and (\ref{e:II53r}), we obtain
\begin{eqnarray}
\langle x(\tau)\rangle_{(\alpha,z)} & = & 
\sqrt{\lfrac{2}{\omega}}|\alpha|\cos{(\omega\tau-
\delta)},\label{e:ssho5}
\\*[1mm]
\langle p(\tau)\rangle_{(\alpha,z)} & = &
-\sqrt{2\omega}|\alpha|\sin{(\omega\tau-\delta)},
\label{e:ssho6}
\end{eqnarray} 
where $\alpha$ and $z$ are defined in Eq. (\ref{e:II30}).  
In addition, according to Eq. (\ref{e:II45r}), we have  
\begin{equation}
|\alpha|^2 = \lfrac{1}{2\omega}(p_o^2+ \omega^2 x_o^2),
~~~\delta = \tan^{-1}{(\lfrac{p_o}{\omega x_o})}.
\label{e:ssho10}
\end{equation}
Note that in this representation, these expectation values are  
independent of the squeeze parameters, $|z|$ and $\theta$.  This  
is because both coherent and squeezed states follow the classical  
motion $x_{cl}(\tau)$ and $p_{cl}(\tau)$.  The differences are that  
the squeezed state is a Gaussian wave packet 
whose width oscillates with time  
and it has a time-dependent uncertainty product, as we come to below. 

With Eqs. (\ref{e:II55r}) and (\ref{e:II58r}), we find that, for
the $(z,\alpha)$-representation,
\begin{eqnarray}
\langle x(\tau)\rangle_{(z,\alpha)} & = & 
2|\alpha|\{\cos{(\omega\tau-\delta)}\cosh{r}\nonumber\\*[ 
1.5mm]
                                    &   & 
~~~~~-\cos{(\omega\tau+\delta-\theta)}\sinh{r}\},
\label{e:ssho15}
\\*[1mm]
\langle p(\tau)\rangle_{(z,\alpha)} & = &
-2\omega|\alpha|\{\sin{(\omega\tau-
\delta)}\cosh{r}\nonumber\\*[1mm]
                                    &   &
~~~~~-\sin{(\omega\tau+\delta-\theta)}\sinh{r}\},
\label{e:ssho16}
\end{eqnarray}
with the connection (\ref{e:II46r}) to the initial position and 
momentum
\begin{eqnarray}
|\alpha|\left[\cos{\delta}\cosh{r}-\cos{(\theta-\delta)}\sinh{r}
\right]
 & = & \sqrt{\lfrac{\omega}{2}}x_0,\nonumber\\*[1mm]
|\alpha|\left[\sin{\delta}\cosh{r}-\sin{(\theta-\delta)}\sinh{r}
\right]
 & = & \sqrt{\lfrac{1}{2\omega}}p_o.\label{e:ssho20}
\end{eqnarray}

\indent From Eq. (\ref{e:ssho20}), we find the identity
\begin{equation}
\lfrac{1}{2\omega}(p_o^2+\omega^2 x_o^2) =  
|\alpha|^2[\cosh{2r} - 
\cos{(\theta-2\delta)}\sinh{2r}].\label{e:ssho24}
\end{equation}
The uncertainty in position and momentum are 
\begin{equation}
(\Delta x)^2 = \lfrac{1}{2\omega}\{\cosh{2r}+
\cos{(2\omega\tau-\theta)}\sinh{2r}\}
\label{e:ssho28}
\end{equation}
and 
\begin{equation}
(\Delta p)^2 = \lfrac{\omega}{2}\{\cosh{2r} + 
\cos{(2\omega\tau-\theta)}\sinh{2r}\},
\label{e:ssho29}
\end{equation}
respectively.  Therefore, the uncertainty product is 
\begin{eqnarray}
(\Delta x)^2(\Delta p)^2 &=& \lfrac{1}{4}[1+ 
\sin^2{(2\omega\tau-\theta)}\sinh^2{2r}]\nonumber
\\*[1mm]
&=& \lfrac{1}{4}\left[1+ 
\frac{\sin^2{(2\omega\tau-\theta)}}{4}
\left(s^2 - \frac{1}{s^2}\right)^2\right],
\label{e:ssho32}
\end{eqnarray}
where \cite{mmn1,mmn2}
\be
s = \exp{r}
\ee
is the ``squeeze parameter."  The uncertainty relation (\ref{e:ssho32}) 
is identical to Eq. (30) in  Ref. {\cite{ntF}} after a suitable choice 
for the phase $\theta$.

For $z$ real and positive, the squeezed-state wave function is 
\begin{equation}
\psi_{ss}(x) = [\pi s^2]^{-1/4}
\exp\left[-\frac{(x-x_o)^2}{2s^2}+ip_ox\right],\label{psiss}
\end{equation}
with $\hbar/m\omega=1$.  For a more general expression for the squeezed 
state wave function, see Ref. {\cite{ntF}.
   
\subsection{Free particle (FP)}

Substituting the real functions (\ref{e:fp4}) into Eqs. 
(\ref{e:II35r}) and (\ref{e:II40r}), we get the expectation values 
for position and momentum in terms of $x_o$ and $p_o$:
\begin{eqnarray}
\langle x(\tau)\rangle & = & x_o + p_o\tau,\label{e:ssfp1}
\\*[1mm]
\langle p(\tau)\rangle & = & p_o.\label{e:ssfp2}
\end{eqnarray}
In the $(\alpha,z)$-representation, we obtain 
\begin{eqnarray}
\langle x(\tau)\rangle_{(\alpha,z)} & = &  
\sqrt{2}|\alpha|[\cos{\delta} + 
\tau\sin{\delta}],\label{e:ssfp5}\\*[1mm]
\langle p(\tau)\rangle_{(\alpha,z)} & = & \sqrt{2}|\alpha|\sin{\delta}. 
\label{e:ssfp6}
\end{eqnarray}
where
\begin{equation}
\sqrt{2}|\alpha|\cos{\delta} = x_o,~~~\sqrt{2}|\alpha|\sin{\delta} =
 p_o,\label{e:ssfp10}
\end{equation}
and
\begin{equation}
|\alpha|^2 = \lfrac{1}{2}(p_o^2+x_o^2),~~~\delta = 
\tan^{-1}{(\lfrac{p_o}{x_o})}.\label{e:ssfp15}
\end{equation}
For the $(z,\alpha)$-representation, we find that 
\begin{eqnarray}
\langle x(\tau)\rangle_{(z,\alpha)} & = & 
\sqrt{2}|\alpha|\{\cos{\delta}\cosh{r}-
\cos{(\theta-\delta)}\sinh{r}\nonumber\\*[1mm]
                                    &   &
~~~~~+\tau [\sin{\delta}\cosh{r}-
\sin{(\theta-\delta)}\sinh{r}]\},\label{e:ssfp20}
\\*[1mm]
\langle p(\tau)\rangle_{(z,\alpha)} & = &
\sqrt{2}|\alpha|\{\sin{\delta}\cosh{r}-
\sin{(\theta-\delta)}\sinh{r}\},\label{e:ssfp21}
\end{eqnarray} 
where  
\begin{eqnarray}
\sqrt{2}|\alpha|[\cos{\delta}\cosh{r}-
\cos{(\theta-\delta)}\sinh{r} & = & 
x_o,\nonumber\\*[1mm]
\sqrt{2}|\alpha|[\sin{\delta}\cosh{r}-
\sin{(\theta-\delta)}\sinh{r} & = & p_o,\label{e:ssfp24}
\end{eqnarray}
and we have the relationship
\begin{equation}
\lfrac{1}{2}(x_o^2+p_o^2) = |\alpha|^2\left[\cosh{2r}-
\cos{(\theta-2\delta)}\sinh{2r}\right].\label{e:ssfp26}
\end{equation}

The uncertainty in position is given by 
\begin{eqnarray}
(\Delta x)^2 & = & \lfrac{1}{2}\left[1+\tau^2\right]\cosh{2r}
\nonumber\\*[1mm]
             &   & ~~~-\lfrac{1}{2}\left\{[1-\tau^2]\cos{\theta}+ 
2\tau\sin{\theta}\right\}\sinh{2r},
\label{e:ssfp30}
\end{eqnarray}
and the uncertainty in momentum is
\begin{equation}
(\Delta p)^2 = \lfrac{1}{2}[\cosh{2r} + 
\cos{\theta}\sinh{2r}].\label{e:ssfp32}
\end{equation}
Therefore, the uncertainty product is
\begin{eqnarray}
(\Delta x)^2(\Delta p)^2 & = & \lfrac{1}{4}\{1 + \tau^2
+[\tau^2\cos{\theta} -\tau\sin{\theta}]\sinh{4r}\} 
\nonumber\\*[1mm]
                         &   & +[\lfrac{1}{2}+\lfrac{3}{2}\tau^2
-\lfrac{1}{2}\left(1-\tau^2\right)\cos{2\theta}-
\tau\sin{2\theta}]\sinh^22{r}.
\label{e:ssfp36}  
\end{eqnarray}
 
\subsection{Linear potential (LP)}

When the interaction is linear, we use the time-dependent  
functions computed in Section 3.3.
Calculating the expectation values of position and momentum 
in terms of the initial position and momentum, we find 
\begin{eqnarray}
\langle x(\tau)\rangle & = & x_o + p_o\tau +\int_0^{\tau}d\rho\,
g(\rho)\rho -\tau\int_0^{\tau}d\rho\,g(\rho), 
\label{e:sslp1}
\\*[1mm]
\langle p(\tau)\rangle & = & p_o -\int_0^{\tau}d\rho\,g(\rho).
\label{e:sslp2}
\end{eqnarray}
For the particular case of $g(\tau)=\kappa/2$, a constant, 
\begin{eqnarray}
\langle x(\tau)\rangle & = & x_o + p_o\tau -\frac{\kappa}{4}\tau^2
\label{e:sslp5}
\\*[1mm]
\langle p(\tau)\rangle & = & p_o -\frac{\kappa}{2}\tau.
\label{e:sslp6}
\end{eqnarray}
To connect to the $(\alpha,z)$-representation, we would have
\begin{eqnarray} 
|\alpha|^2 = \lfrac{1}{2}\left[(x_0-g_o)^2+p_o^2\right],
\nonumber\\*[1mm]
\delta = \tan^{-1}{\left({{p_o}\over{x_o-g_o}}\right)}.
\label{e:sslp10}
\end{eqnarray}
For the $(z,\alpha)$-representation, Eqs. (\ref{e:II46r}) become 
\begin{eqnarray}
|\alpha|\left[\cos{\delta}\cosh{r}-\cos{(\theta-\delta)}\sinh{r}
\right] & = & \sqrt{\lfrac{1}{2}}\left(x_o-g_o\right),\nonumber
\\*[1mm]
|\alpha|\left[\sin{\delta}\cosh{r}-\sin{(\theta-\delta)}\sinh{r}
\right] & = & \sqrt{\lfrac{1}{2}}p_o. \label{e:sslp15}
\end{eqnarray}
We can combine these two equations into one to yield
\begin{equation}
|\alpha|\left[\cosh{2r}-cos{(\theta-2\delta)}\sinh{2r}\right] 
=\lfrac{1}{2}\left(p_o^2+x_o^2-2g_ox_o+g_o^2\right).
\label{e:sslp16}
\end{equation}
To obtain the corresponding equations for constant $g(\tau)$, replace 
$g_o$ by $\kappa/2$ in each of the above equations.

For the $(\alpha,z)$-representation, the corresponding expectation  
values for $x$ and $p$ are
\begin{eqnarray}
\langle x(\tau)\rangle_{(\alpha,z)} & = &  
\sqrt{2}|\alpha|\left[\cos{\delta} + \tau\sin{\delta}\right] 
+\int_0^{\tau}d\rho\,g(\rho)\rho+g_o
-\tau\int_0^{\tau}d\rho\,g(\rho),\label{e:sslp20}
\\*[1mm] 
\langle p(\tau)\rangle_{(\alpha,z)} & = & \sqrt{2}|\alpha|
\sin{\delta}-\int_0^{\tau}d\rho\,g(\rho).
\label{e:sslp21} 
\end{eqnarray} 
When $g(\tau)=\kappa/2$, we find from the previous two equations  
that 
\begin{eqnarray}
\langle x(\tau)\rangle_{(\alpha,z)} & = &  
\sqrt{2}|\alpha|\left[\cos{\delta} + \tau\sin{\delta}\right] 
-\frac{\kappa}{4}\tau^2+\frac{\kappa}{2},\label{e:sslp22}
\\*[1mm] 
\langle p(\tau)\rangle_{(\alpha,z)} & = & \sqrt{2}|\alpha|
\sin{\delta}-\frac{\kappa}{2}\tau.\label{e:sslp23} 
\end{eqnarray}
For the $(z,\alpha)$-representation, we see that 
\begin{eqnarray}
\langle x(\tau)\rangle_{(z,\alpha)} & = & 
\sqrt{2}|\alpha|\left\{\cos{\delta}\cosh{r}-
\cos{(\theta-\delta)}\sinh{r}\right.\nonumber\\
                                    &   &
~~~~~+\left.\tau \left[\sin{\delta}\cosh{r}-
\sin{(\theta-\delta)}\sinh{r}\right]\right\}\nonumber\\
                                    &   &
~~~~~+\int_0^{\tau}d\rho\,g(\rho)\rho+g_o
-\tau\int_0^{\tau}d\rho\,g(\rho),
\label{e:sslp25}
\\*[1.5mm]
\langle p(\tau)\rangle_{(z,\alpha)} & = &
\sqrt{2}|\alpha|\left\{\sin{\delta}\cosh{r}-
\sin{(\theta-\delta)}\sinh{r}\right\}
-\int_0^{\tau}d\rho\,g(\rho).
\label{e:sslp26}
\end{eqnarray} 
When $g(\tau)=\kappa/2$,  
\begin{eqnarray}
\langle x(\tau)\rangle_{(z,\alpha)} & = & 
\sqrt{2}|\alpha|\left\{\cos{\delta}\cosh{r}-
\cos{(\theta-\delta)}\sinh{r}\right.\nonumber\\
                                    &   &
~~~~~~+\left.\tau \left[\sin{\delta}\cosh{r}-
\sin{(\theta-\delta)}\sinh{r}\right]\right\}\nonumber\\
                                    &   &
~~~~~~-\frac{\kappa}{4}\tau^2+\frac{\kappa}{2},
\label{e:sslp28}
\\*[1.5mm]
\langle p(\tau)\rangle_{(z,\alpha)} & = &
\sqrt{2}|\alpha|\left\{\sin{\delta}\cosh{r}-
\sin{(\theta-\delta)}\sinh{r}\right\}
-\frac{\kappa}{2}\tau.
\label{e:sslp29}
\end{eqnarray}

The uncertainties in position and momentum are given by Eq.  
(\ref{e:ssfp30}) and (\ref{e:ssfp32}), respectively.  Therefore, 
the uncertainty product (\ref{e:ssfp36}) is still valid for 
a system with a linear interaction since it is independent of 
$g(\tau)$. 

\subsection{Driven Harmonic Oscillator (HDO)}

Referring to the results of Section 3.4 and Eqs. 
(\ref{e:II35r}) and (\ref{e:II40r}), the expectation 
values for position and momentum are presented 
below.

\noindent (a) In the $(x_o,p_o)$-representation:
\begin{eqnarray}
\langle x(\tau)\rangle & = & \lfrac{1}{\omega}
\left(p_o\sin{\omega\tau}+\omega x_o\cos{\omega\tau}\right)
\nonumber\\
                       &   & +\lfrac{1}{\omega}
\left\{\cos{\omega\tau}\int_0^{\tau}d\rho\,
g(\rho)\sin{\omega\tau}-\sin{\omega\tau}\int_0^{\tau}d\rho\,
g(\rho)\cos{\omega\tau}\right\},\label{e:ssdo1}\\*[1mm]
\langle p(\tau)\rangle & = & p_o\cos{\omega\tau}
-\omega x_o\sin{\omega\tau}\nonumber\\
                       &   & -\sin{\omega\tau}
\int_0^{\tau}d\rho\,g(\rho)\sin{\omega\tau}
-\cos{\omega\tau}\int_0^{\tau}d\rho\,g(\rho)\cos{\omega\tau},
\label{e:ssdo2}
\end{eqnarray}
where, for the $(\alpha,z)$-representation 
\begin{eqnarray}
 & |\alpha|^2 = \lfrac{1}{2\omega}\left(p_o^2+\omega^2x_o^2
+g_ox_o+\lfrac{1}{\omega^2}g_o^2\right), & \nonumber\\*[1mm]
 & \delta = \tan^{-1}{{{\omega p_o}\over{\omega^2+g_o}}}.
\label{e:ssdo5}
\end{eqnarray}
For the $(z,\alpha)$-representation, the connecting formulas 
are
\begin{eqnarray}
 & |\alpha|\left[\cos{\delta}\cosh{r}-\cos{(\theta-\delta)}
\sinh{r}\right] = \sqrt{\lfrac{1}{2}}\left(\sqrt{\omega}x_o
+\lfrac{1}{\omega^{3/2}}g_o\right), & \nonumber\\*[1mm]
 & |\alpha|\left[\sin{\delta}\cosh{r}-\sin{(\theta-\delta)}
\sinh{r}\right] = \sqrt{\lfrac{1}{2\omega}}p_o.\label{e:ssdo8}
\end{eqnarray}

\noindent (b) In the $(\alpha,z)$-representation:
\begin{eqnarray} 
\langle x(\tau)\rangle_{(\alpha,z)} & = &
\sqrt{\lfrac{2}{\omega}}|\alpha|\cos{(\omega\tau-\delta)} +
(\chi_1{\cal C}_2 - \chi_2{\cal C}_1),\label{e:ssdo12}
\\*[1mm]  
\langle p(\tau)\rangle_{(\alpha,z)} & = &
-\sqrt{2\omega}|\alpha|\sin{(\omega\tau-\delta)} + 
(\dot{\chi}_1{\cal C}_2 - \dot{\chi}_2{\cal C}_1),
\label{e:ssdo13} 
\end{eqnarray}

\noindent (c) In the $(z,\alpha)$-representation:
\begin{eqnarray} 
\langle x(\tau)\rangle_{(z,\alpha)} & = &
2|\alpha|\{\cos{(\omega\tau-\delta)}\cosh{r}
\nonumber\\
                                    &   & 
~~~~~-\cos{(\omega\tau+\delta-\theta)}\sinh{r}\} +  
(\chi_1{\cal
C}_2 - \chi_2{\cal C}_1),\label{e:ssdo16}
\\*[1mm] 
\langle p(\tau)\rangle_{(z,\alpha)} & = &
-2\omega|\alpha|\left\{\sin{(\omega\tau-\delta)}\cosh{r}\right.
\nonumber\\
                                    &   &
~~~~~\left.-\sin{(\omega\tau+\delta-\theta)}\sinh{r}\right\} +
\dot{\chi}_1{\cal C}_2 - \dot{\chi}_2{\cal C}_1.
\label{e:ssdo17}
\end{eqnarray} 

In each of (b) and (c) above, we have 
\begin{eqnarray}
\chi_1{\cal C}_2-\chi_2{\cal C}_2 & = & 
\lfrac{1}{\omega}\{\cos{\omega\tau}
\int_{\tau_o}^{\tau}d\rho\,g(\rho)\sin{\omega\rho}\nonumber\\
                                  &   &
~~~~~~-\sin{\omega\tau}
\int_{\tau_o}^{\tau}d\rho\,g(\rho)\cos{\omega\rho}\},
\label{e:ssdo20}
\end{eqnarray} 
and
\begin{eqnarray}
\dot{\chi}_1{\cal C}_2-\dot{\chi}_2{\cal C}_2 & = & 
-\sin{\omega\tau}\int_{\tau_o}^{\tau}d\rho\,g(\rho)
\sin{\omega\rho}\nonumber\\
                                        &   &
~~~~~~-\cos{\omega\tau}
\int_{\tau_o}^{\tau}ds\,g(\rho)\cos{\omega\rho}.
\label{e:ssdo21}
\end{eqnarray}

When  $g(\tau)=\kappa/2$, explicit expressions of 
expectation values for $x$ and $p$ can be obtained.  From
Eqs. (\ref{e:ssdo1}) and {\ref{e:ssdo2}), we find  
\begin{eqnarray}
\langle x(\tau)\rangle & = & \lfrac{1}{\omega}
\left(p_o\sin{\omega\tau}+\omega x_o\cos{\omega\tau}\right)
+\lfrac{\kappa}{2\omega^2}\left(\cos{\omega\tau}-1\right),
\label{e:ssdo24}\\*[1mm]
\langle p(\tau)\rangle & = & p_o\cos{\omega\tau}-\omega x_o
\sin{\omega\tau}-\lfrac{\kappa}{2\omega}\sin{\omega\tau}.
\label{e:ssdo25}
\end{eqnarray}
For the $(\alpha,z)$-representation, applying Eqs. 
(\ref{e:ssdo12}) and (\ref{e:ssdo13}), we find that 
\begin{eqnarray}
\langle x(\tau)\rangle_{(\alpha,z)} & = & 
\sqrt{\lfrac{2}{\omega}}
|\alpha|\cos{(\omega\tau-\delta)}-\lfrac{\kappa}{2\omega^2},
\label{e:ssdo28}\\*[1mm]
\langle p(\tau)\rangle_{(\alpha,z)} & = & -\sqrt{2\omega}
|\alpha|\sin{(\omega\tau-\delta)}.\label{e:ssdo29}
\end{eqnarray}

Making use of Eqs. (\ref{e:ssdo16}) and (\ref{e:ssdo17}), 
we obtain for the $(z,\alpha)$-representation, 
\begin{eqnarray}
\langle x\rangle & = & \sqrt{\lfrac{2}{\omega}}\left\{
\cos{(\omega\tau-\delta)}\cosh{r}
-\sin{(\omega\tau+\delta-\theta)}\sinh{r}\right\}
-\lfrac{\kappa}{2\omega^2},\label{e:ssdo32}\\*[1mm]
\langle p\rangle & = & -\sqrt{2\omega}|\alpha|
\left\{\sin{(\omega\tau-\delta)}\cosh{r}
-\sin{(\omega\tau+\delta-\theta)}\sinh{r}\right\}.\label{ssdo33}
\end{eqnarray}
The connecting formulas for this case can be obtained from 
Eqs. (\ref{e:ssdo5}) and (\ref{e:ssdo8}) by substituting 
$g_o=k/2$.

Expressions for the uncertainties in position and momentum,  
(\ref{e:ssho28}) and (\ref{e:ssho29}), respectively, derived 
for HO remain valid here.  As a consequence, the 
uncertainty product (\ref{e:ssho32}) holds for DHO.   
\vskip .2cm

\subsection{Repulsive oscillator (RO)}

Referring to Section 3.5, for the time-dependent functions for 
the repulsive oscillator, we find the the expectation values in the 
$(x_o,p_o)$-representation are
\begin{eqnarray}
\langle x(\tau)\rangle & = &  
\lfrac{1}{\Omega}\left[p_o\sinh{\Omega\tau} 
+ \Omega  
x_o\cosh{\Omega\tau}\right],\label{e:ssro1}
\\*[1mm]
\langle p(\tau)\rangle & = & p_o\cosh{\Omega\tau} 
+ \Omega x_o\sinh{\Omega\tau}.\label{e:ssro2}
\end{eqnarray}

For the $(\alpha,z)$-representation, we obtain
\begin{eqnarray}
\langle x(\tau)\rangle_{(\alpha,z)} & = & 
\sqrt{\lfrac{2}{\Omega}}|\alpha|[\cosh{\Omega\tau}
\cos{\delta} +
\sinh{\Omega\tau}\sin{\delta}],\label{e:ssro5}
\\*[1mm]
\langle p(\tau)\rangle_{(\alpha,z)} & = &
\sqrt{2\Omega}|\alpha|[\sinh{\Omega\tau}\cos{\delta} +
\cosh{\Omega\tau}\sin{\delta}],\label{e:ssro6} 
\end{eqnarray} 
where we have  
\begin{equation} 
x_o = \sqrt{\lfrac{2}{\Omega}}|\alpha|\cos{\delta},~~~p_o
= \sqrt{2\Omega}|\alpha|\sin{\delta},\label{e:ssro10} 
\end{equation}
and 
\begin{equation}
|\alpha|^2 = \lfrac{1}{2\Omega}(p_o^2 + \Omega^2  
x_o^2).\label{e:ssro12}
\end{equation}

In the $(z,\alpha)$-representation, we see that
\begin{eqnarray}
\langle x(\tau)\rangle_{(z,\alpha)} & = & 
\sqrt{\lfrac{2}{\Omega}}|\alpha|\{[\cos{\delta}\cosh{r}\nonumber\\
                                   &   & 
~~~~~-\cos{(\theta-\delta)}\sinh{r}]
\cosh{\Omega\tau}\nonumber\\
                                   &   & 
~~+[\sin{\delta}\cosh{r}-\sin{(\theta-\delta)}
\sinh{r}]\sinh{\Omega\tau},\label{e:ssro16}
\\*[1mm]
\langle p(\tau)\rangle_{(z,\alpha)} & = &
\sqrt{2\Omega}|\alpha|\{[\cos{\delta}\cosh{r}
\nonumber\\*[1.5mm]
                                   &   & 
~~~~~-\cos{(\theta-\delta)}\sinh{r}]
\sinh{\Omega\tau}\nonumber\\
                                   &   & 
~~+[\sin{\delta}\cosh{r}-\sin{(\theta-\delta)}\cosh{\Omega\tau)}\},
\label{e:ssro24}
\end{eqnarray} 
where we have 
\begin{eqnarray}
x_o & = & \sqrt{\lfrac{2}{\Omega}}|\alpha|[\cos{\delta}\cosh{|z|} - 
\cos{(\theta-\delta)}\sinh{|z|}],\nonumber\\*[1mm]
p_o & = & \sqrt{2\Omega}|\alpha|[\sin{\delta}\cosh{|z|}-
\sin{(\theta-\delta)}\sinh{|z|}].\label{e:ssro25}
\end{eqnarray}
In addition, we have the identity
\begin{equation}
\lfrac{1}{2\Omega}(p_o^2+\Omega^2 x_o^2) = 
|\alpha|^2\left[\cosh{2r}-\cos{(\theta-\delta)}\sinh{2r}
\right].\label{e:ssro30}
\end{equation}

We obtain the uncertainty product directly from Eq. (93) of paper I.  
We have
\begin{eqnarray}
(\Delta x)^2(\Delta p)^2 & = & \lfrac{1}{4}\left(1+\sinh^2{2\Omega\tau}
\right) -\lfrac{1}{4}\sinh^2{2\Omega\tau}\sin{\theta}\sinh{4r},
\nonumber\\
                         &   & +\lfrac{1}{8}
\left\{1+3\sinh^2{2\Omega\tau}+\cosh^2{2\Omega\tau}\cos{2\theta}
\right\}\sinh^2{2r}.\label{e:ssro34}
\end{eqnarray}
Initially, the Gaussian wave packet describing this state satisfies 
the minimum uncertainty condition, but spreads out over time.

\vspace{.2cm}

%*****************************

\section{Discussion}

All quantum systems described by a Schr\"odinger equation 
(\ref{e:I2}) with potential (\ref{e:I3}) 
have isomorphic symmetry algebras, designated by 
$({\cal SA})^c_1$ or its oscillator subalgebra 
$os(1)=\{{\cal M}_3,{\cal J}_{\pm},I\}$.  This 
isomorphism means that for each such system it is possible to 
construct a complete set of eigenstates of the operator 
${\cal M}_3$ and the 
Casimir operator, ${\bf C}$, of $os(1)$.  These states form a 
representation space for $os(1)$.  They are also eigenstates of 
the number operator, ${\cal J}_+{\cal J}_-$, 
constructed from the ladder operators, ${\cal J}_{\pm}$, of $os(1)$.  
This is a consequence of the relationship  between the operators ${\cal M}_3$ 
and ${\cal K}_3$ (Eq. (28) of paper I).  
Only for HO do these states correspond to 
energy eigenstates.  

In Sec. 3 of paper I, we showed that the extremal 
state is a 
Gaussian function.  For all of the systems discussed in this paper, the 
limit as $\tau\rightarrow 0$ of ${\cal K}_3$ is an time-independent 
oscillator Hamiltonian (HO, FP, RO) or a time-independent driven 
oscillator Hamiltonian (LP, DHO).  (See Sec. 3.3 and 3.4 of this paper.)  
Therefore, each system has effectively been transformed by the 
${\cal R}$-separable coordinates, $(\zeta,\eta)$ (Sec. 3, paper I), 
into a time-independent oscillator or a driven oscillator.

In the genral treatment of 
paper I, we computed squeezed-state wave functions for both 
the $(\alpha,z)$- 
and $(z,\alpha)$-representations.  
[See Eqs. (63) and (64) of I.]  
Each of them were written 
as expansions in eigenstates of the number operator.  We may think  
of these expansions as representing transformed Gaussian functions 
{\cite{mmn96}}.  
Expectation values for position and momentum, uncertainties in 
position and momentum, and their uncertainty product were 
derived.  

In the $(\alpha,z)$-representation, according to Eqs. 
(\ref{e:II50r}) and (\ref{e:II53r}), $\langle x\rangle$ and 
$\langle p\rangle$ depend only on $\alpha$ and 
not on $z$.  Since $\alpha$ is fixed by the initial position and 
momentum, $z$ is free to vary. 
[See Eq. (\ref{e:II45r}).]  However, 
in the $(z,\alpha)$-representation, according to Eqs. (\ref{e:II55r}) 
and (\ref{e:II58r}), the expectation values depend upon all four 
parameters $|\alpha|$, $\delta$, r, and $\theta$.  
From Eq. (\ref{e:II46r}), we can determine any two of these 
in terms of the initial position and initial momentum.  
A third way of expressing the expectation values of $x$ and $p$ is 
in terms of initial position and momentum.  [See Eqs. (\ref{e:II35r}) 
and (\ref{e:II40r}).]
This way is independent of the complex parameters $\alpha$ and $z$, 
and therefore is identical for both representations.  Only then are 
the relationships between the four parameters, $\alpha$, $\delta$, 
$r$, and $\theta$, and the initial position and momentum sensitive 
to which representation we are using.

The expectation values of $x$ 
and $p$ satisfy the classical equations of motion and 
describe classical trajectories in phase space: for HO  
the trajectory is an ellipse; for FP it is a straight line; for LP 
it is a parabola; for DHO it is  
a displaced ellipse, and for RO it is a hyperbola.

The uncertainty products do not depend on $\alpha$, but do 
depend on $z$.  Also, they  
are independent of representation.  For HO and DHO the 
time-dependence of the uncertainty product is 
linked to the squeeze parameter, $z$.  If $z=0$, then the uncertainty 
product is minimized.  When $z\ne 0$, there is an 
oscillation in the uncertainty product, subject to 
$(\Delta x)^2(\Delta p)^2\ge \lfrac{1}{4}$.  For the other three 
systems, the uncertainty product increases with time, starting 
from a state of minimum uncertainty.  The 
Gaussian wave packet for these three systems will eventually dissipate.      

%********************************

\section*{Acknowledgements}

MMN acknowledges the support of the United States Department of 
Energy.  DRT acknowledges
a grant from the Natural Sciences and Engineering Research Council 
of Canada.

%************************


\newpage

\begin{thebibliography}{99}

\bibitem{drt1} D. R. Truax, J. Math. Phys. {\bf 22}, 1959 (1981).

\bibitem{drt2} D. R. Truax, J. Math. Phys. {\bf 23}, 43 (1982).

\bibitem{gt} S. Gee and D.R. Truax, Phys. Rev. A {\bf 29}, 1627  
(1984).

\bibitem{kt1} A. Kalivoda and D.R. Truax, to be published.

\bibitem{ntF} M.M. Nieto and D.R. Truax, Forschritte der Physik, 
in press.

\bibitem{mmn1} M. M. Nieto, in {\it Frontiers of Nonequilibrium  
Statistical 
Physics}, edited by G. T. Moore and M. O. Scully (Plenum, New  
York, 1986).

\bibitem{mmn2} M. M. Nieto, Quantum Opt. {\bf 6}, 9 (1994).  The  
large round 
brackets in Eq. (5) should be squared.  See. Eq. (4.11) of Ref.  
\cite{mmn1}.

\bibitem{mmn96} M. M. Nieto, Quantum Semiclass. Opt., in press.

\end{thebibliography}
\end{document}